\documentclass[iop]{emulateapj}
\usepackage{epsfig}
\usepackage{apjfonts}
\usepackage{amsmath,color,bm}
\usepackage{booktabs}
\usepackage[breaklinks,colorlinks,urlcolor=blue,citecolor=blue,linkcolor=blue]{hyperref}
\shorttitle{Lensing of Continuous Waves from Spinning Neutron Stars}
\shortauthors{Basak et al}

\usepackage{natbib}
\usepackage{nameref}
\usepackage{graphicx}
\usepackage{graphics}
\usepackage[space]{grffile}
\usepackage{latexsym}
\usepackage{amsfonts,amsmath,amssymb}
\usepackage{url}
\usepackage[utf8]{inputenc}
\usepackage{fancyref}
\usepackage{hyperref}
\usepackage{xcolor}
\usepackage[normalem]{ulem}

\newcommand{\red}[1]{{\textcolor{black} {{#1}}}}

\begin{document}

\title{Prospects for the observation of continuous gravitational waves from spinning neutron stars lensed by the galactic supermassive black hole}

\author{Soummyadip Basak$^{1*}$}
\author{Aditya Kumar Sharma$^{1*}$}
\author{Shasvath J. Kapadia$^1$}
\author{Parameswaran Ajith$^{1,2}$}
\affiliation{$^1$~International Centre for Theoretical Sciences, Tata Institute of Fundamental Research, Bangalore 560089, India}
\affiliation{$^2$~Canadian Institute for Advanced Research, CIFAR Azrieli Global Scholar, MaRS Centre, West Tower, 661 University Ave, Toronto, ON M5G 1M1, Canada}
\altaffiliation{$^*$ Equal contribution from both authors.}

\begin{abstract}

We study the prospects of detecting continuous gravitational waves (CGWs) from spinning neutron stars (NSs), gravitationally lensed by the galactic supermassive black hole. Assuming various astrophysically motivated spatial distributions of galactic NSs, we find that CGW signals from a few ($\sim 0-6$) neutron stars should be strongly lensed. Lensing will produce two copies of the signal (with time delays of seconds to minutes) that will interfere with each other. The relative motion of the NS with respect to the lensing optical axis will change the interference pattern, which will help us to identify a lensed signal. Accounting for the magnifications and time delays of the lensed signals, we investigate their detectability by ground-based detectors. Modelling the spin distribution of NSs based on that of known pulsars and assuming an ellipticity of $\epsilon = 10^{-7}$, lensed CGWs are unlikely to be detectable by LIGO and Virgo in realistic searches involving $\mathcal{O}(10^{12})$ templates. However, third generation detectors have a \red{$\sim 2-51\%$} probability of detecting at least one lensed CGW signal.  For an ellipticity of $\epsilon = 10^{-8}$, the detection probability reduces to \red{$\sim 0-18 \, \% $}. Though rare, such an observation will enable interesting probes of the supermassive black hole and its environment. 

\end{abstract}

\section{Introduction}\label{sec:introduction}
LIGO and Virgo detectors \citep{aLIGO, Virgo} have detected $\sim 100$ transient gravitational-wave (GW) signals during their first three observing runs \citep{gwtc-2, abbott2021gwtc}, most of which are consistent with GWs produced by coalescing binary black holes (BBHs). GWs from merging binary neutron stars \citep{GW170817-detection, GW190425-detection} and neutron star-black hole binaries \citep{NSBH-Discovery} have also been observed. These detections have afforded a plethora of scientific riches, including an unprecedented probe of the population of compact binaries \citep{O3a-Rates}, a distance-ladder-independent measurement of the Hubble constant \citep{O2-Hubble}, as well as some of the most stringent tests of Einstein's general theory of relativity in the strong-field regime \citep{O3a-tgr}.

Although there has so far been no confident detection of the gravitational lensing of GWs \citep{O3a-Lensing} \footnote{Note, however, that some tantalizing candidates of lensed GW signals have been proposed; see, e.g., ~\cite{Dai:2020tpj}.}, there is a growing consensus in the literature that lensed GWs from merging BBHs are likely to be detected in the upcoming observing runs of LIGO, Virgo and KAGRA \citep{KAGRA:2020tym} \citep[see, e.g.:][]{Ng2018}. Observations of such events will provide additional insights into various aspects of astrophysics, cosmology and fundamental physics. Apart from being the very first detection of gravitational lensing involving a new messenger, they will enable accurate localisation of the host galaxy of the merger~\citep{Hannuksela:2020xor}, provide unique constraints on the constituents of dark matter \citep{Jung:2017flg,Urrutia:2021qak,Basak:2021ten}, on models of the populations of galaxies and galaxy clusters \citep{Smith2019}, as well as on alternative theories of gravity \citep{Fan2017,Ezquiaga2020, Goyal2021}.

While the list of detections of transient GWs has been growing from one observing run to the next \citep{gwtc-1, gwtc-2}, and is expected to grow even more drastically in the future \citep{abbott2020prospects}, continuous GWs (CGWs) remain undetected \citep{O2-CW-AllSky, O3a-CW-AllSky}. Rapidly spinning, non-axisymmetric neutron stars (NSs) in our galaxy are expected to produce CGWs potentially observable by ground-based detectors~\citep[see, e.g.:][]{Bonazzola1996}. 

In this {\it letter}, we explore the prospects of observing the gravitational lensing of CGWs from spinning NSs by the galactic supermassive black hole (SMBH) \citep{schodel2002star,  ghez2003first}. Such an observation will provide potentially powerful probes of the properties of the astrophysical source as well as lens. Focusing exclusively on strong lensing, we expect the CGWs to be lensed if the source NS resides within the Einstein angle of the lens. Assuming the SMBH to be a point mass lens, strong lensing will produce two copies of a  CGWs, with a time delay between them~\footnote{CGWs from rapidly spinning NSs, with spin frequencies spanning $\sim 100-1000$ Hz, have wavelengths that are $\mathcal{O}(10^3-10^4)$ times smaller than the Schwarzschild radius of the galactic SMBH. The geometric optics approximation therefore holds for the lensing scenario considered here.}. The copies will have differing amplitudes, although their time-dependent phase will be identical. The image waveforms will show-up in the detector as one superposed CGW, whose amplitude will depend on the magnifications of the images as well as the time-delay between the two copies of the signal at the detector.

The number of NSs that are expected to lie within the Einstein angle of the SMBH will depend on the (poorly known) spatial distribution of NSs in the galaxy. We consider various astrophysically motivated distributions presented in the literature, and evaluate the distribution of the number of NSs that fall within the Einstein angle, assuming a total of $10^9$ NSs in the galaxy~\citep{Treves:1999ne}. We find that up to 6 NSs will be within the Einstein angle of the SMBH, so that their CGWs, if detected, will be strongly lensed. 

We further assess the detectability of these signals by third generation (3G) GW detector network consisting of two Cosmic Explorers~\citep{Evans:2021gyd} and one Einstein Telescope~\citep{Punturo_2010}, incorporating the effects of lensing magnification and time delays. The detectability, characterized by the signal-to-noise ratio (S/N), is proportional to their amplitude, as well as the square root of the observation time \citep{Jaranowski1998}. The amplitude, in turn, is proportional to the ellipticity, the moment of inertia and the square of the spin frequency of the NS (apart from  extrinsic parameters such as the location and orientation). 

We assume an ellipticity of $10^{-7}$, which is an order of magnitude smaller than the best upper limits obtained from a directed search for NSs in the galactic center, for a fiducial moment of inertia of $10^{38}~\mathrm{kg\,m^2}$~\citep{KAGRA:2022osp}. Spin frequencies are drawn from the spin distribution of known pulsars~\citep{ATNF}. The signal amplitude is averaged over the inclination angle of the NS rotation axis with respect to the line of sight, over the angle between the rotation axis and the axis of symmetry, as well as the polarization angle.  Using a single template search (i.e., assuming that the source parameters are known a priori) the probability of detecting at least one lensed CGW signal is $\sim \red{0-15 \%~ (2-53\%)}$ in LIGO-Virgo (third generation detectors).  For a more realistic, directed search towards the galactic centre using $\sim 10^{12}$ templates~\citep{aasi2013directed}, the corresponding probability is $\sim \red{0-2\% ~ (2-51\%)}$. Note that the ellipticity of most neutron stars could be much lower. For a more conservative assumption of $\epsilon =10^{-8}$, LIGO-Virgo detectors are unlikely to detect any lensed signals. In 3G detectors the detection probability is \red{$\sim  1-36\%~ (\sim 0-18\%)$}  for a single template search (a directed search involving $10^{12}$ templates). If the ellipticity is lower than $10^{-8}$, the detection probability will be even smaller.

A possible detection will enable very interesting probes of the physics and astrophysics of the source as well as lens. The lensed CGW signal will contain imprints of the properties of the SMBH, such as its mass and spin, enabling an independent measurement of these properties. Such an observation might also enable us to constrain the presence of additional hairs of the black hole, thus probing the true nature of the supermassive compact object at the galactic center. In addition, stars and stellar-mass compact objects in the galactic centre can cause additional microlensing effects on the CGW signal~\citep[e.g.,][]{Liao:2019aqq,Suvorov:2021uvd}. This will potentially allow us to probe the poorly understood astrophysical environment of the galactic centre. Any proper motion of the NS will also leave an imprint in the CGW signal. 

The rest of the paper is organized as follows. Section ~\ref{sec:lensing} briefly introduces gravitational lensing by a point mass lens. Section~\ref{sec:distributions} describes the spatial distributions of NSs assumed, as well as the resulting estimate on the number NSs  strongly lensed by the galactic SMBH.  Section~\ref{sec:detectability} delineates the calculation of the S/N and provides the (S/N-threshold-dependent) probability of detecting a lensed CGW in the 3G era. Section ~\ref{sec:conclusion} summarizes the paper, discusses a potential means of identifying lensed CGW candidates, and the possible astrophysical measurements that can be performed from such an observation.

\section{Gravitational lensing by a point-mass lens}\label{sec:lensing}

\begin{figure}[tb]
	\includegraphics[width=0.9\linewidth]{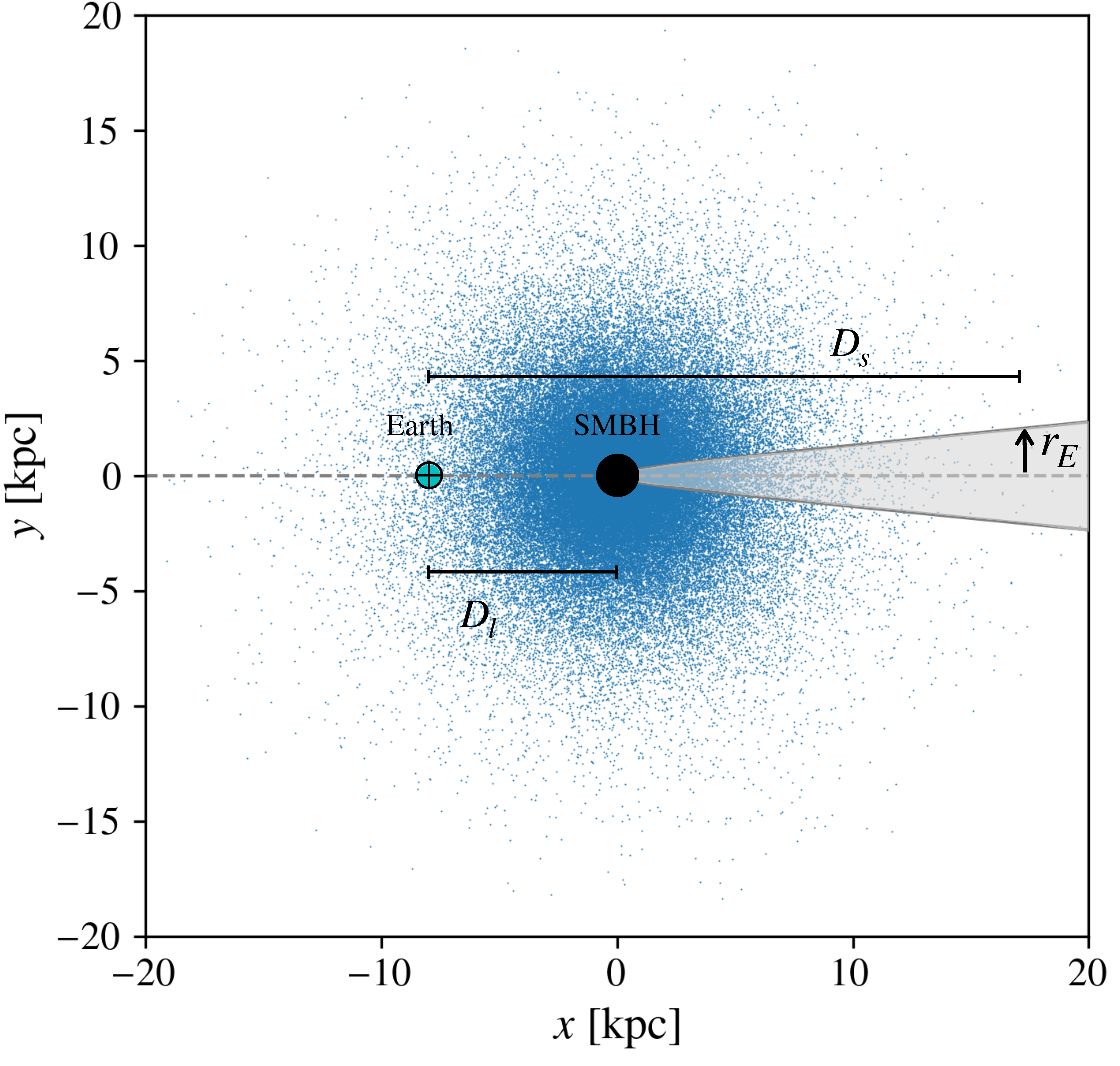}
	\caption{Schematic diagram of the distribution of neutron stars (blue dots) projected onto our galactic plane. The Einstein cone of the central SMBH is shown by the gray region (highly exaggerated). The neutron stars within the Einstein cone will be strongly lensed by the SMBH.}
	\label{fig:schematic}
\end{figure}

The strong lensing of GWs, in the geometric optics limit, is identical to that of the lensing of electromagnetic waves, and applies in general to null geodesics \citep[see, e.g:][]{dodelson2017}. Thus, as with the gravitational lensing of light, the fundamental equation that governs strong lensing of GWs is the so-called lens equation that relates the source location $\vec{\beta}$, with the image location $\vec{\theta}$, via a deflection angle $\vec{\alpha}(\vec{\theta})$
\begin{equation}
	\vec{\beta} = \vec{\theta} - \vec{\alpha}(\vec{\theta}).
\end{equation}
Note that $\vec{\beta}, \vec{\theta}$ are angles measured with respect to the line connecting the earth and the lens, called the optical axis. The  deflection angle $\vec{\alpha}(\vec{\theta})$ depends on the relative locations of the earth, the lens and the source, as well as the gravitational potential of the lens. For a point-mass lens with mass $M_L$, 
\begin{equation}
	\vec{\alpha} = \frac{\theta^2_E}{\theta^2}\vec{\theta}, 
\end{equation}
where $\theta_E$ is the Einstein angle (Einstein radius). In terms of the Schwarzschild radius $R_s = {2GM_L}/{c^2}$ of the lens, this angle can be written as:
\begin{equation}\label{eq:EinsteinAngle}
	\theta_E = \sqrt{\frac{2R_sD_{{SL}}}{D_SD_L}}.
\end{equation}
Here, $D_S$, $D_L$ are the distances (from earth) to the source and the lens, respectively, and $D_{SL} = D_S-D_L$ \footnote{Since the distances considered in this work are galactic, cosmological effects are negligible. These distances can therefore be approximated to be Euclidean.}. Multiple images are produced when the source is within the Einstein angle of the lens. This is a conservative assumption, as in the case of a point mass lens, multiple images can be produced even when the source is outside the Einstein radius. Solving the lens equation with the deflection angle for a point mass lens yields two images at locations:
\begin{equation}
	\theta_{\pm} = \frac{\beta}{2}\left[1 \pm \sqrt{1 + \frac{4\theta_E^2}{\beta^2}} \right ].
\end{equation}
The magnifications of the images can be acquired from the inverse of the determinant of the Jacobian transformation matrix between $\vec{\beta}$ and $\vec{\theta}$:
\begin{equation}
	\mu = { \det \left (\frac{\partial \vec{\beta}}{\partial\vec{\theta}} \right) }^{-1}.
	\label{eq:magnification}
\end{equation}
For a point mass lens, the magnifications of the two images reduces to:
\begin{equation}
	\mu_{\pm} = \left[1 - \left(\frac{\theta_{E}}{\theta_{\pm}}\right)^4 \right]^{-1}.
\end{equation}
The time delay in the arrival of the two images at the earth has two contributing pieces. There is a geometric time delay due to the different paths travelled by the rays pertaining to each of the images. There is also a Shapiro time delay, caused by the general relativistic time dilation suffered by the rays when they venture into the vicinity of the lens. 
For a point mass lens, the total time delay for sources that lie within the Einstein angle is to a good approximation given by:
\begin{equation}
	\Delta t \simeq 2\frac{D_\mathrm{L} D_\mathrm{S}}{c D_{\mathrm{SL}}}\theta_E\beta. 
	\label{eq:timedelay}
\end{equation}

\section{Number of strongly lensed neutron stars}\label{sec:distributions}

Investigating the prospects of detecting lensed CGWs involves counting the expected number of NSs within the Einstein angle of the SMBH and assessing the detectability of CGWs produced by them. This requires assumptions on the total number of NSs in our galaxy and their spatial distribution. While it is generally believed that $\sim 10^9$ NSs reside in the our galaxy, only $\sim 10^3$  have been detected so far through electromagnetic observations. Thus, little is known about the statistical properties of galactic neutron stars, including their spatial distribution.

We consider three types of spatial distribution of NSs in the galaxy. One assumes that NSs have the same distribution of stars in the young galactic disk. Following~\cite{Paczynski:1990}, we write the probability distributions in galactocentric cylindrical coordinates system $(R, \phi, z)$, where the $z$ axis corresponds to the rotation axis of the Milky Way, as: 
\begin{eqnarray}
	\frac{dP}{dR}  &=& a_R \, \frac{R}{R^2_0}\exp\left(\frac{-R}{R_0} \right), 
	\label{eq:progenitor_model_R} \\ 
	\frac{dP}{dz} &=& \frac{1}{2 z_0} \exp\left(\frac{-|z|}{z_0} \right), 
	\label{eq:progenitor_model_z} 
\end{eqnarray}
where $R_0 = 4.5$ kpc and $z_0 = 0.07$ kpc are scaling constants. We call this the ``progenitor'' model. 

This will be a good approximation of the spatial distribution of the NSs if their natal kicks are small or when the NSs are young. However, the distribution of NSs can differ from that of stars depending on the NS birth velocities, which remains largely uncertain. To mimic the effect of natal kicks on the spatial distribution of NSs, some authors have considered different choices of $z_0$ in Eq.\eqref{eq:progenitor_model_z}. For e.g., \cite{Reed:2021scb} uses a range of $z_0$ values out of which we choose four different values ($z_0 = 0.1, 0.2, 0.5, 1$ kpc) in Eq.\eqref{eq:progenitor_model_z} along with a Gaussian-like distribution in $R$. 
\begin{equation}
	\frac{dP}{dR} = \frac{R}{\sigma_{\mathrm{R}}^2} \exp \left( \frac{-R^2}{2\sigma_{\mathrm{R}}^2}  \right), 
	\label{eq:Reed_model_R}
\end{equation}
where $\sigma_{\mathrm{R}} = 5~\mathrm{kpc}$. Several studies also have evolved  populations of NSs in the galactic potential by considering different models of the birth velocity to predict the expected distribution of NSs in the present epoch. \cite{Sartore:2010}  assumes that the NSs are born in the galaxy with a constant birth rate, at locations given by the progenitor distribution presented in Eqs.\eqref{eq:progenitor_model_R}-\eqref{eq:progenitor_model_z}. They evolved this distribution under several different assumptions on their birth velocities (indicated by A, B, C, D and E), and two different models of the galactic potential (models with and without a ``*''). By fitting their simulation data, \cite{Sartore:2010} presented the following fitting functions:
\begin{eqnarray}
	\frac{dP}{dR} & \propto & R ~ \exp \left( a_0 + a_1 R + a_2 R^2 + a_3 R^3 + a_4 R^4 \right) \\ 
	\frac{dP}{dz}  & \propto & \frac{1}{b_0 b_1^z + b_2}, 
	\label{eq:sartore_model}
\end{eqnarray}
where the fitting coefficients are tabulated in Tables A.1 and A.2 of ~\cite{Sartore:2010}. 

For all models, we finally construct the 3-dimensional distribution 
\begin{equation}
	\frac{dP}{dR d\phi dz}  = C \, \frac{dP}{dR} \, \frac{dP}{d\phi}  \, \frac{dP}{dz} 
\end{equation}
where all models assume axial symmetry around the rotation axis of the galaxy ($dP/d\phi = 1/2 \pi$). The distributions in \cite{Paczynski:1990} and \cite{Reed:2021scb} are already normalised; hence $C = 1$. For \cite{Sartore:2010} models, the normalisation constant $C$ is determined by the condition that a certain fraction of the NSs presently resides in the disk of the galaxy. 
\begin{equation}
	\int_{R=0}^{R_\mathrm{disk}} dR \int_{\phi=0}^{2\pi} d\phi \int_{z = -z_\mathrm{disk}}^{z_\mathrm{disk}} dz \frac{dP}{dR d\phi dz}  = f_\mathrm{disk}.
\end{equation}
Above, $R_\mathrm{disk} = 20$~kpc and $z_\mathrm{disk} = 0.2$~kpc, while $f_\mathrm{disk}$ is given in Table 4 of \cite{Sartore:2010}. 

\begin{figure}[tb]
	\includegraphics[width=\linewidth]{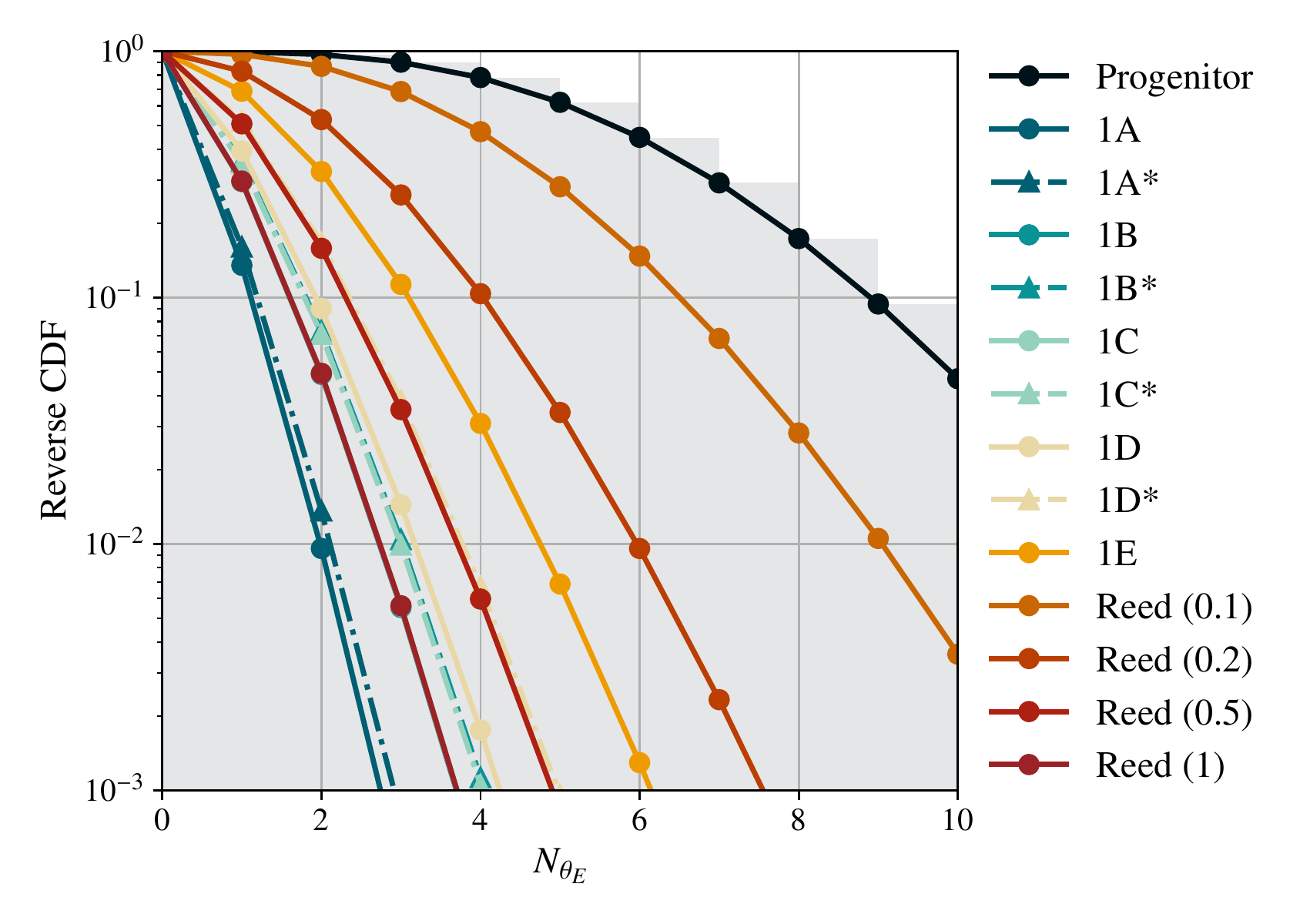}
	\caption{Cumulative distribution of the expected number of NSs within the Einstein angle of the galactic SMBH as predicted by different models of the NS spatial distribution and galactic potential. ``Progenitor'' model assumes that the spatial distribution of NSs follows that of the stars in the galaxy~\citep{Paczynski:1990}. Models 1A, 1B, 1C, 1D and 1E are predicted by~\cite{Sartore:2010} assuming different models of NS birth velocities. The dashed curves correspond to the models 1A*, 1B*, 1C* and 1D* of \cite{Sartore:2010} which assumes the same NS birth velocities but a different model for the galactic potential. For the \cite{Reed:2021scb} models, we consider different values of $z_0$ (shown in brackets, in kpc). The probability of at least one NS being inside the lensing cone, $P(N_{\theta_E} \geq 1) $, is $\sim 0.1 - 1$, depending on the model.}
	\label{fig:N_thetaE_disst}
\end{figure}

In order to find the average number of NSs that would be strongly lensed (producing multiple images), we integrate the probability $dP/dRd\phi dz$ over a cone-like region around the optical axis with a radius of $r_E \equiv \theta_E D_s$ (shaded region Fig.~\ref{fig:schematic}), and multiply it with the total expected number of NSs in the galaxy ($N \sim 10^9$). 
\begin{equation}
	\bar{N}_{\theta_E} = N \int_\mathrm{lensing~cone} dR d\phi dz \frac{dP}{dR d\phi dz} . 
\end{equation}
Depending on the distribution $\bar{N}_{\theta_E}$ varies from 0.1 to 5.6. 
Assuming no spatial clustering of NSs, the actual number, $N_{\theta_E}$, of NSs that will be strongly lensed by the SMBH will be distributed according to a Poisson distribution with mean $\bar{N}_{\theta_E}$. Figure~\ref{fig:N_thetaE_disst} shows the distribution of $N_{\theta_E}$ for various models of the NS spatial distribution and galactic potential. We see that the probability of at least one NS being inside the lensing cone is significant $\sim 0.1 - 1$,  depending on the model). 

\section{Lensed continuous GWs and their detectability}\label{sec:detectability}

\begin{figure}[tb]
	\includegraphics[width=\linewidth]{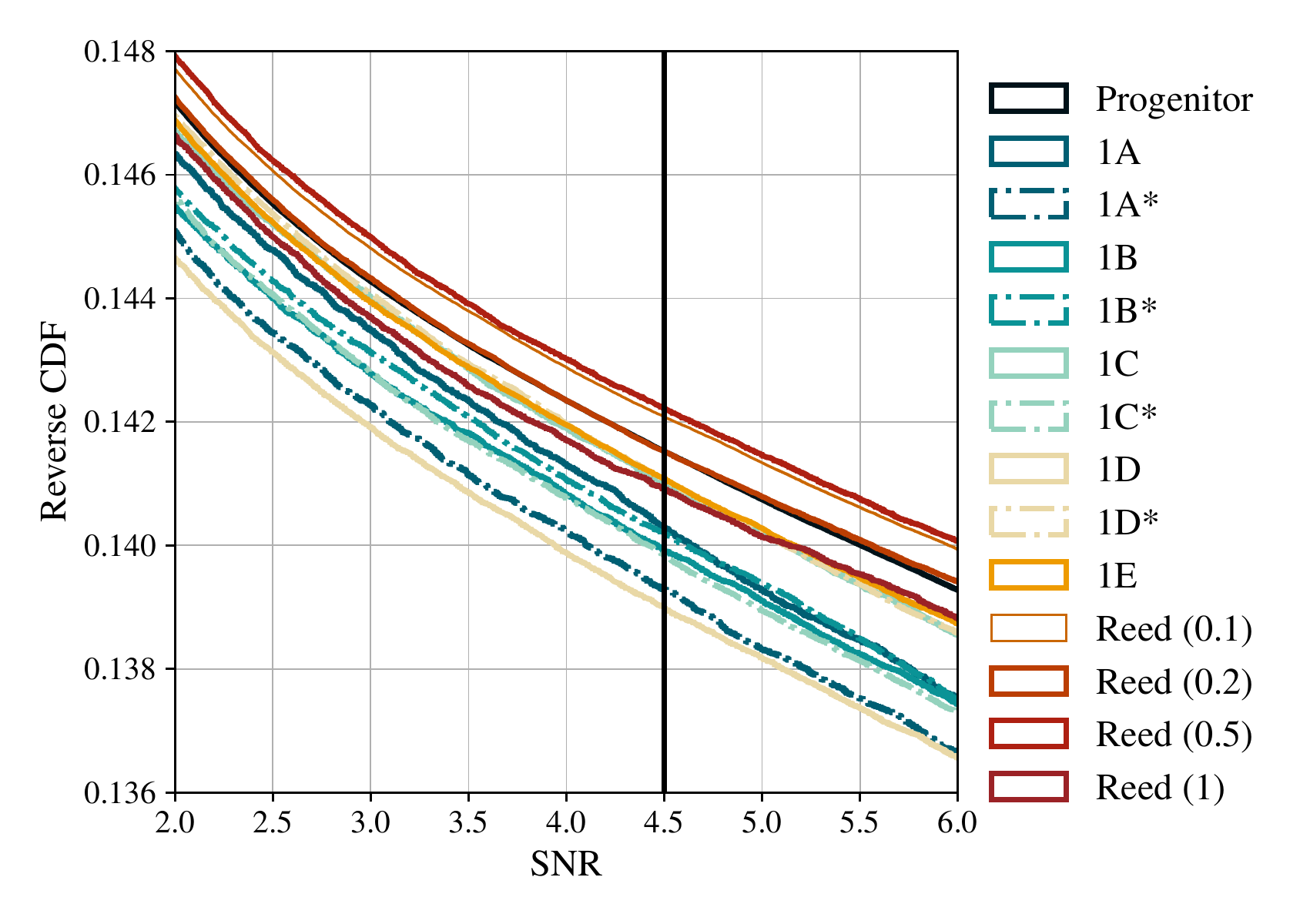}
	\caption{Cumulative distribution of the S/Ns of the lensed NSs for  different spatial distributions (same as Fig.\ref{fig:N_thetaE_disst}). The S/N threshold of 4.5 corresponding to a false alarm probability of 1\% and false dismissal probability of 10\% using a single-template search is shown by the vertical line.}
	\label{fig:dist_mu_td}
\end{figure}

In order to assess the detectability of lensed CGWs, we need to compute their S/N at the detector. We model the NS as a triaxial ellipsoid \citep[see, e.g:][]{andersson2019}. Its ellipticity is defined in terms of the moments of inertia around the rotation axis ($I$) and in the plane perpendicular to the principal axis ($I_1, I_2$):
\begin{equation}
	\epsilon = \frac{\left | I_1 - I_2 \right |}{I}.
\end{equation}
If the rotation axis of the spinning NS does not align with the principal axis, the resulting time varying mass quadrupole moment will produce GWs whose amplitude is proportional to $\epsilon$, $I$, as well as the squared rotation frequency $f_{\star}$:
\begin{equation}
	h_0 = \frac{16\pi^2G}{c^4}\frac{\epsilon \, I \, f_{\star}^2}{r},
\end{equation}
where $r$ is the distance to the NS. Furthermore, the frequency of the GWs emanated depends on the mechanism that produces the mass quadrupole moment. 
In general, CGWs are generated at the first and second harmonic of the rotation frequency $f_{\star}$. We only consider the second harmonic ($ f = 2 f_{\star}$)  in this work, because we find that the amplitude of the first harmonic is always lower than that of the second harmonic, which determines the detectability of the signal. 

The corresponding GW polarizations are given by:
\begin{eqnarray}
	h_{}^{+}(t) &=& A_{}^+ \, h_0 \, \cos\left(2\pi f_{}t + \varphi \right), \\
	h_{}^{\times}(t) &=& A_{}^{\times} \, h_0  \, \sin\left(2\pi f_{}t + \varphi \right),
\end{eqnarray}
where $A^{+} = \sin \chi \frac{1 + \cos^2 \iota}{2} $,  $A^{\times} = \sin \chi \cos \iota$. Here, $\iota$ is the inclination angle between the rotation axis and the line of sight, $\chi$ is the angle between the rotation axis and the principal axis (called wobble angle), and $\varphi$ is a constant phase offset.

The measured GW strain at a detector depends on its response to the GWs. This response, characterized by the time-dependent antenna pattern functions $F^{+}(t), F^{\times}(t)$, depends on the relative orientation and location of the detector with respect to the location of the source:
\begin{equation}
	h_{}(t) = F^{+}(t)h^{+}_{}(t) + F^{\times}(t)h_{}^{\times}(t).
\end{equation}
CGWs lensed by the galactic SMBH (modelled as a point-mass lens with mass $M_L$), will produce exactly two images with magnifications $\mu_{\pm}$ and time-delay $\Delta t$. The resulting strain measured at the detector will therefore be a superposition between the two copies of CGWs:
%
%
%
\begin{eqnarray}\label{eq:h-int}
	h_{\mathrm{tot}}(t) &=&  \sqrt{\mu^{\mathrm{int}}}\left[ F^{+}(t)h^{+}_{}(t) + F^{\times}(t)h_{}^{\times}(t)\right ].
\end{eqnarray}
Here, $\mu^{\mathrm{int}}$ is an amplification factor that results from the interference of the two lensed signals, and is given by \footnote{Note that in addition to this amplification, lensing will add a constant phase $\varphi^\mathrm{int}$ to the signal that depends on the magnifications of the images, and the time delay. This can be absorbed in to the phase constant $\varphi$. However, proper motion of the source with respect to the optical axis will make $\mu^{\mathrm{int}}$  and $\varphi^\mathrm{int}$ time-dependent, introducing amplitude and phase modulations in the lensed signal. This will help us distinguish lensed and unlensed signals; see Section~\ref{sec:conclusion}.}
\begin{equation}\label{eq:mag-int}
	\mu^{\mathrm{int}}_{} = \left| \mu_{+} \right | + \left | \mu_{-} \right | + 2\sqrt{\left | \mu_{+}\mu_{-} \right |}\cos(2\pi f_{}\Delta t).
\end{equation}
In order to assess the detectability of such signals, we evaluate an averaged S/N $\rho$ (which depends on an averaged $h_{\mathrm{tot}}^2$), where the average is taken over the period of rotation of the NS (for $h_{}^{+,\times}$), the sidereal day (for $F^{+, \times}$), as well as the inclination angle ($\iota$), polarization angle ($\psi$), and wobble angle ($\chi$).
\begin{equation}
	\rho = \left[\frac{\langle \left (h_{}^{\mathrm{tot}} \right)^2 \rangle (f_{})T_{\mathrm{obs}}}{S_n(f_{})}\right]^{1/2}.
\end{equation}
Here, $S_n(f)$ is the detector's noise power spectral density (PSD), and $T_{\mathrm{obs}}$ is the observation time. 

In order to estimate the detectability of strongly lensed NSs, we simulated populations of NSs within the lensing cone, distributed according to different models. We computed the lensing magnifications and time delays for each NS using Eqs.~\eqref{eq:magnification} and \eqref{eq:timedelay}. We evaluate the single-detector S/N of the superposed images after accounting for interference (cf. Eqs.~\ref{eq:h-int}, \ref{eq:mag-int}), averaged over rotation period of the NS, the wobble angle of the NS, the time-varying antenna patterns of the detector across a sidereal day, and the inclination angle of the rotation axis of the NS with respect to the line of sight.  We set the sky location of the lensed sources to coincide with the location of SMBH (Sagittarius A*), which is a good approximation given that the Einstein angle extends to within an arc-second centered at that location. We assume an ellipticity of $\epsilon = 10^{-7}$, and frequencies drawn from the frequency distribution of pulsars from the ATNF catalog~\citep{ATNF}.  The network S/N is the quadratic sum of the individual detector S/Ns. (see. Fig.~\ref{fig:dist_mu_td}). 

\begin{figure}[tb]
	\includegraphics[width=0.5\textwidth]{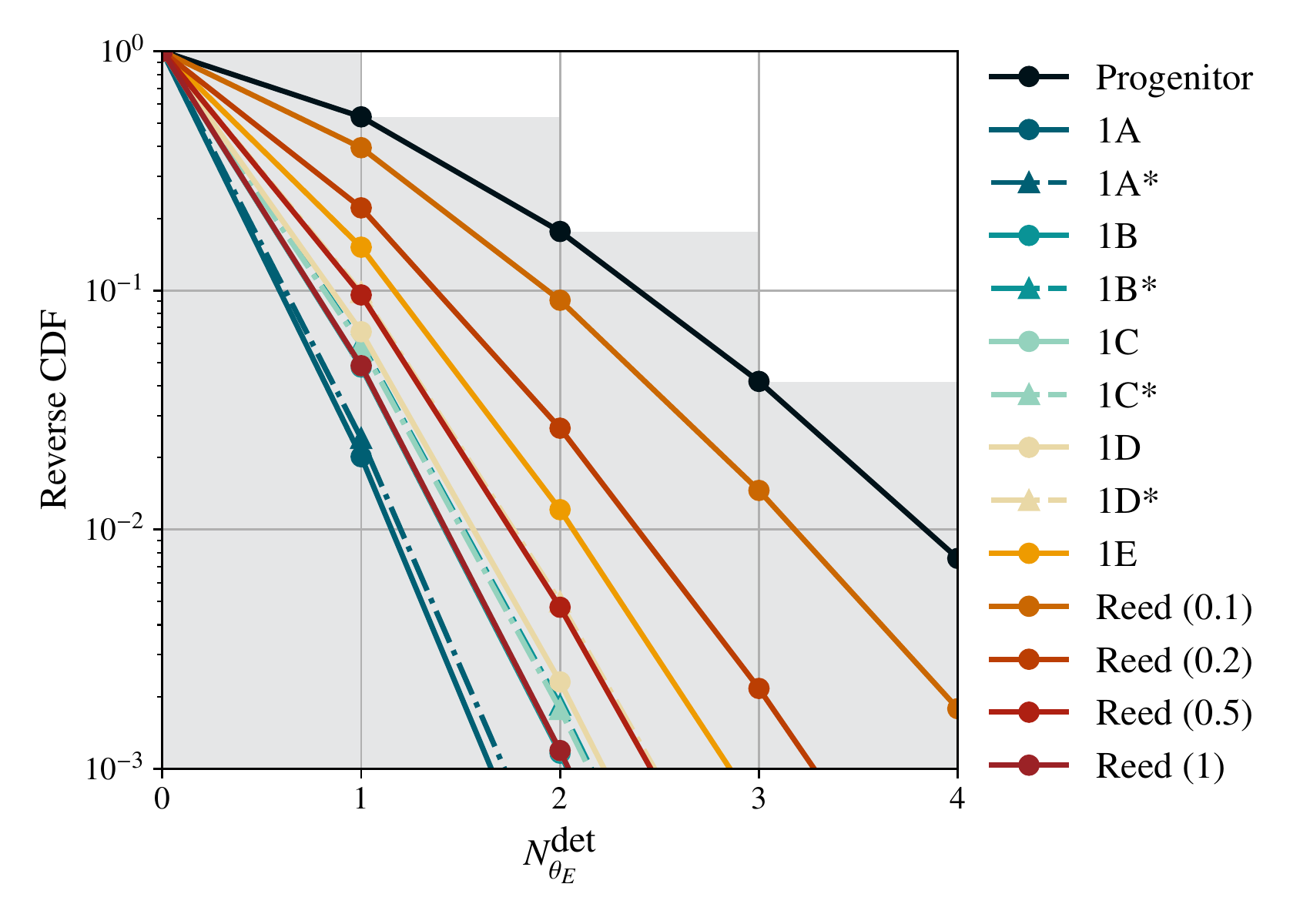}
	\caption{Cumulative distribution of the detectable number of strongly lensed events by 3G detectors with S/N threshold of 4.5. The probability of at least one NS being inside the lensing cone, $P(N_{\theta_E}^\mathrm{det} \geq 1) $, is $\red{\sim 2-53\%}$, depending on the model.  The gray histogram shows the results computed using a simulation using the progenitor model, while the different lines are analytical calculations using Poisson distributions. 
	}
	\label{fig:N_thetaE_det_dist}
\end{figure}

The fraction of detectable NSs within the Einstein angle is 
\begin{equation}
	N_{\theta_{E}}^\mathrm{det} = N_{\theta_{E}} ~ \alpha(\rho_\mathrm{thresh}), ~~~ \alpha(\rho_\mathrm{thresh}) = \int_{\rho_\mathrm{thresh}}^{\infty} \frac{dP}{d\rho} d\rho, 
\end{equation}
where the S/N distribution ${dP}/{d\rho}$ for each model is estimated from simulations. Figure~\ref{fig:N_thetaE_det_dist} shows the distribution of detectable number of NSs within the Einstein cone. Depending on the model and the assumptions of the NS properties, the probability of detecting at least one strongly lensed NS is $\red{\sim 2-53\%}$~\footnote{We assume a three-detector network consisting of two CE detectors and one ET. The expected PSDs are generated from the ``optimal'' curves presented in Fig.\,2 of~\cite{Hall_2019}.}. This is assuming an S/N threshold of 4.5 which corresponds to a false alarm probability (FAP) of 1 $\%$ and a false dismissal probability (FDP) of 10 $\%$ for a single template search.

We also investigated the detection probability of lensed NSs in the fifth observing run (O5) of LIGO, Virgo and KAGRA~\citep{observer_summary}, and find that the range of detection probabilities is $\red{\sim 0-15\%}$~\footnote{We assume a five-detector network involving three LIGO detectors (including LIGO-India), Virgo and KAGRA. The expected PSDs are generated using \textsc{aLIGOAPlusDesignSensitivityT1800042, AdvVirgo, KAGRALateSensitivityT1600593} functions of the PyCBC PSD package~\citep{pycbcpsd}.}. However, a realistic search for such signals would require a bank of templates, since the intrinsic parameters of the NSs are not known a priori. Assuming a directed search towards the galactic centre involving $\sim 10^{12}$ templates \citep{aasi2013directed}, the SNR threshold corresponding to the FAP and FDP mentioned above, but now also accounting for the trials factor due to the  template bank, becomes $\simeq 9.6$. With this threshold, the detection probability in O5 drops to $ \red{\sim 0-2\%}$. In 3G detectors, this probability continues to be non-trivial, with a range of \red{$\sim 2 - 51 \%$}.  If we make a more pessimistic assumption of ellipticity $\epsilon =10^{-8}$, LIGO-Virgo detectors are unlikely to detect any lensed signals. In 3G detectors the detection probability is \red{$\sim  1-36\%$}  for a single template search and \red{$\sim 0-18\%$} for a directed search involving $10^{12}$ templates. Smaller values of $\epsilon$ will reduce the detection probability further.

These estimates are consistent with the non-detection of CGWs by the directed searches towards the galactic centre using LIGO-Virgo data from the third observing run~\citep{KAGRA:2022osp}. Using the spatial and frequency distribution models that we employed to study the detectability of {lensed} signals, we estimate the detection probability of (all) CGWs to be $\sim 0-2\%$ with $\epsilon = 10^{-7}$ and coherent integration time of 1 yr~\footnote{Here we assume that the search is directed towards NSs located in a cone that has its apex on the Earth and has a base radius equal to the Einstein radius of the SMBH for a source at $D_{SL} = 15~ \mathrm{kpc}$.}. For the coherent integration times of a few hrs employed in \cite{KAGRA:2022osp}, the expected detection probability is almost zero. 



\section{Discussion}\label{sec:conclusion}

In this \emph{Letter}, we explored the possibility of detecting CGWs from spinning neutron stars strongly lensed by the galactic SMBH. Treating the SMBH as a point-mass lens, we consider a source to be lensed if it lies within the Einstein angle of the SMBH. To assess the prospects of detecting such lensed CGWs, we considered several spatial distributions of NSs presented in the literature. We find that up to 6 out of $10^9$ NSs lie within the Einstein angle. 
Accounting for the lensing magnification and time delays as well as the resulting interference between the two images, we evaluate the detectability of such sources. 

Unlike the lensing of GW transients such as  compact binary coalescences, which gives temporally resolved copies of signals whose morphology can be compared to determine whether they are lensed, lensed CGWs would show-up in the  data as a single, interfered signal. If lensing introduces a constant time delay, the interfered signal would be indistinguishable from an unlensed CGW with the same amplitude, except for a constant phase shift. However, if the relative transverse motion between the NS and the lensing optical axis (axis connecting the Earth and the SMBH) is sufficiently large, the time delay $\Delta t$ between the lensed copies of the CGW itself become a function of time. This will result in the modulation of the amplitude and phase of the lensed CGW signals, rendering them identifiable. 

Generically, we expect some relative motion between the NS and the optical axis. This could be  due to the proper motion of the NS in the galaxy (e.g., due to the natal kicks; $v \sim 100$ km/s), due to the motion of the Earth around the Sun  ($v \sim 30$ km/s), or due to the differential motion of the solar system in the galactic potential ($v \sim 10$ km/s). A simple, back-of-the envelope calculation can give an estimate of the degree of this modulation. From Eq.\eqref{eq:timedelay}, the accumulated change in the lensing time delay over an observational time $T$ can be estimated as $\simeq 2\frac{D_\mathrm{L} D_\mathrm{S}}{c \, D_{\mathrm{SL}}}\theta_E \, \frac{d\beta}{dt} ~ T \simeq 2\frac{D_\mathrm{L}}{c \, D_{\mathrm{SL}}}\theta_E \, v \, T$. This can cause several modulation cycles in the amplitude and phase of the CGW signal over the course of a year, helping us to identify lensed signals. 

Lensed CGWs, if detected, would enable unique probes of astrophysics and gravity. For example, the lensing time delay and hence the amplitude and phase modulation of the lensed CGW signal depends on the mass of the SMBH.  Such an observation would be a unique new way of measuring the mass of the galactic SMBH. In addition, compact objects and stars in the galactic centre could produce additional microlensing effects on the GW signal, which are potentially measurable~\citep{Liao:2019aqq,Suvorov:2021uvd,2020PhRvD.101b4039M}. This would be a powerful means of probing the astrophysical environment of the galactic centre. Unlike electromagnetic radiation GWs do not suffer from extinction, and can potentially provide an uncontaminated picture. Lensed CGW signals can, in principle, contain signatures of additional properties of the SMBH, such as its spin angular momentum~\citep{Gralla:2019drh}, and more speculatively, other possible ``hairs''~\citep{Islam:2021dyk}. They will also allow us to measure the proper motion of the NS. Lensing of CGWs harbour a rich and complex phenomenology, which we plan to explore in upcoming work.


\paragraph{Acknowledgments:}

We thank David Keitel for reviewing the manuscript and providing useful comments. We acknowledge support of the Department of Atomic Energy, Government of India, under project no. RTI4001.  SJK’s work was supported by a grant from the Simons Foundation (677895,  R.G.) to the International Centre for Theoretical Sciences, Tata Institute of Fundamental Research (ICTS-TIFR).  PA’s research was supported by the Canadian Institute for Advanced Research through the CIFAR Azrieli Global Scholars program. Computations were performed with the aid of the Alice computing cluster at ICTS-TIFR.
\bibliography{references}

\begin{thebibliography}{}
\expandafter\ifx\csname natexlab\endcsname\relax\def\natexlab#1{#1}\fi
\providecommand{\url}[1]{\href{#1}{#1}}
\providecommand{\dodoi}[1]{doi:~\href{http://doi.org/#1}{\nolinkurl{#1}}}
\providecommand{\doeprint}[1]{\href{http://ascl.net/#1}{\nolinkurl{http://ascl.net/#1}}}
\providecommand{\doarXiv}[1]{\href{https://arxiv.org/abs/#1}{\nolinkurl{https://arxiv.org/abs/#1}}}

\bibitem[{Aasi {et~al.}(2013)}]{aasi2013directed}
Aasi, J., {et~al.} 2013, Phys. Rev. D, 88, 102002,
  \dodoi{10.1103/PhysRevD.88.102002}

\bibitem[{{Aasi} {et~al.}(2015)}]{aLIGO}
{Aasi}, J., {et~al.} 2015, Classical and Quantum Gravity, 32, 074001,
  \dodoi{10.1088/0264-9381/32/7/074001}

\bibitem[{{Abbott} {et~al.}(2017{\natexlab{a}})}]{GW170817-detection}
{Abbott}, B.~P., {et~al.} 2017{\natexlab{a}}, \prl, 119, 161101,
  \dodoi{10.1103/PhysRevLett.119.161101}

\bibitem[{{Abbott} {et~al.}(2017{\natexlab{b}})}]{O2-Hubble}
---. 2017{\natexlab{b}}, \nat, 551, 85, \dodoi{10.1038/nature24471}

\bibitem[{Abbott {et~al.}(2018)}]{observer_summary}
Abbott, B.~P., {et~al.} 2018, Living Rev. Rel., 21, 3,
  \dodoi{10.1007/s41114-020-00026-9}

\bibitem[{{Abbott} {et~al.}(2019{\natexlab{a}})}]{gwtc-1}
{Abbott}, B.~P., {et~al.} 2019{\natexlab{a}}, Physical Review X, 9, 031040,
  \dodoi{10.1103/PhysRevX.9.031040}

\bibitem[{{Abbott} {et~al.}(2019{\natexlab{b}})}]{O2-CW-AllSky}
---. 2019{\natexlab{b}}, \prd, 100, 024004, \dodoi{10.1103/PhysRevD.100.024004}

\bibitem[{{Abbott} {et~al.}(2020)}]{GW190425-detection}
---. 2020, \apjl, 892, L3, \dodoi{10.3847/2041-8213/ab75f5}

\bibitem[{Abbott {et~al.}(2020)Abbott, Abbott, Abbott, Abraham, Acernese,
  Ackley, Adams, Adya, Affeldt, Agathos, {et~al.}}]{abbott2020prospects}
Abbott, B.~P., Abbott, R., Abbott, T., {et~al.} 2020, Living reviews in
  relativity, 23, 1

\bibitem[{{Abbott} {et~al.}(2020{\natexlab{a}})}]{gwtc-2}
{Abbott}, R., {et~al.} 2020{\natexlab{a}}, arXiv e-prints, arXiv:2010.14527.
\newblock \doarXiv{2010.14527}

\bibitem[{{Abbott} {et~al.}(2020{\natexlab{b}})}]{O3a-tgr}
---. 2020{\natexlab{b}}, arXiv e-prints, arXiv:2010.14529.
\newblock \doarXiv{2010.14529}

\bibitem[{Abbott {et~al.}(2021)Abbott, Abbott, Acernese, Ackley, Adams,
  Adhikari, Adhikari, Adya, Affeldt, Agarwal, {et~al.}}]{abbott2021gwtc}
Abbott, R., Abbott, T., Acernese, F., {et~al.} 2021, arXiv preprint
  arXiv:2111.03606

\bibitem[{{Abbott} {et~al.}(2021{\natexlab{a}})}]{NSBH-Discovery}
{Abbott}, R., {et~al.} 2021{\natexlab{a}}, \apjl, 915, L5,
  \dodoi{10.3847/2041-8213/ac082e}

\bibitem[{{Abbott} {et~al.}(2021{\natexlab{b}})}]{O3a-Rates}
---. 2021{\natexlab{b}}, \apjl, 913, L7, \dodoi{10.3847/2041-8213/abe949}

\bibitem[{{Abbott} {et~al.}(2021{\natexlab{c}})}]{O3a-Lensing}
---. 2021{\natexlab{c}}, arXiv e-prints, arXiv:2105.06384.
\newblock \doarXiv{2105.06384}

\bibitem[{{Abbott} {et~al.}(2021{\natexlab{d}})}]{O3a-CW-AllSky}
---. 2021{\natexlab{d}}, \prd, 103, 064017, \dodoi{10.1103/PhysRevD.103.064017}

\bibitem[{Abbott {et~al.}(2022)}]{KAGRA:2022osp}
Abbott, R., {et~al.} 2022, Phys. Rev. D, 106, 042003,
  \dodoi{10.1103/PhysRevD.106.042003}

\bibitem[{{Acernese} {et~al.}(2015)}]{Virgo}
{Acernese}, F., {et~al.} 2015, Classical and Quantum Gravity, 32, 024001,
  \dodoi{10.1088/0264-9381/32/2/024001}

\bibitem[{Akutsu {et~al.}(2021)}]{KAGRA:2020tym}
Akutsu, T., {et~al.} 2021, PTEP, 2021, 05A101, \dodoi{10.1093/ptep/ptaa125}

\bibitem[{Andersson(2019)}]{andersson2019}
Andersson, N. 2019, Gravitational-wave astronomy: Exploring the dark side of
  the Universe (Oxford Graduate Texts)

\bibitem[{Basak {et~al.}(2022)Basak, Ganguly, Haris, Kapadia, Mehta, \&
  Ajith}]{Basak:2021ten}
Basak, S., Ganguly, A., Haris, K., {et~al.} 2022, Astrophys. J. Lett., 926,
  L28, \dodoi{10.3847/2041-8213/ac4dfa}

\bibitem[{{Bonazzola} \& {Gourgoulhon}(1996)}]{Bonazzola1996}
{Bonazzola}, S., \& {Gourgoulhon}, E. 1996, \aap, 312, 675.
\newblock \doarXiv{astro-ph/9602107}

\bibitem[{Dai {et~al.}(2020)Dai, Zackay, Venumadhav, Roulet, \&
  Zaldarriaga}]{Dai:2020tpj}
Dai, L., Zackay, B., Venumadhav, T., Roulet, J., \& Zaldarriaga, M. 2020.
\newblock \doarXiv{2007.12709}

\bibitem[{Dodelson(2017)}]{dodelson2017}
Dodelson, S. 2017, Gravitational lensing (Cambridge University Press)

\bibitem[{Evans {et~al.}(2021)}]{Evans:2021gyd}
Evans, M., {et~al.} 2021.
\newblock \doarXiv{2109.09882}

\bibitem[{{Ezquiaga} \& {Zumalac{\'a}rregui}(2020)}]{Ezquiaga2020}
{Ezquiaga}, J.~M., \& {Zumalac{\'a}rregui}, M. 2020, \prd, 102, 124048,
  \dodoi{10.1103/PhysRevD.102.124048}

\bibitem[{{Fan} {et~al.}(2017){Fan}, {Liao}, {Biesiada},
  {Pi{\'o}rkowska-Kurpas}, \& {Zhu}}]{Fan2017}
{Fan}, X.-L., {Liao}, K., {Biesiada}, M., {Pi{\'o}rkowska-Kurpas}, A., \&
  {Zhu}, Z.-H. 2017, \prl, 118, 091102, \dodoi{10.1103/PhysRevLett.118.091102}

\bibitem[{Ghez {et~al.}(2003)Ghez, Duch{\^e}ne, Matthews, Hornstein, Tanner,
  Larkin, Morris, Becklin, Salim, Kremenek, {et~al.}}]{ghez2003first}
Ghez, A., Duch{\^e}ne, G., Matthews, K., {et~al.} 2003, The Astrophysical
  Journal Letters, 586, L127

\bibitem[{{Goyal} {et~al.}(2021){Goyal}, {Haris}, {Mehta}, \&
  {Ajith}}]{Goyal2021}
{Goyal}, S., {Haris}, K., {Mehta}, A.~K., \& {Ajith}, P. 2021, \prd, 103,
  024038, \dodoi{10.1103/PhysRevD.103.024038}

\bibitem[{Gralla \& Lupsasca(2020)}]{Gralla:2019drh}
Gralla, S.~E., \& Lupsasca, A. 2020, Phys. Rev. D, 101, 044031,
  \dodoi{10.1103/PhysRevD.101.044031}

\bibitem[{Hall \& Evans(2019)}]{Hall_2019}
Hall, E.~D., \& Evans, M. 2019, Classical and Quantum Gravity, 36, 225002,
  \dodoi{10.1088/1361-6382/ab41d6}

\bibitem[{Hannuksela {et~al.}(2020)Hannuksela, Collett, \c{C}al\i{}\c{s}kan, \&
  Li}]{Hannuksela:2020xor}
Hannuksela, O.~A., Collett, T.~E., \c{C}al\i{}\c{s}kan, M., \& Li, T. G.~F.
  2020, Mon. Not. Roy. Astron. Soc., 498, 3395, \dodoi{10.1093/mnras/staa2577}

\bibitem[{Islam \& Ghosh(2021)}]{Islam:2021dyk}
Islam, S.~U., \& Ghosh, S.~G. 2021, Phys. Rev. D, 103, 124052,
  \dodoi{10.1103/PhysRevD.103.124052}

\bibitem[{Jaranowski {et~al.}(1998)Jaranowski, Kr\'olak, \&
  Schutz}]{Jaranowski1998}
Jaranowski, P., Kr\'olak, A., \& Schutz, B.~F. 1998, Phys. Rev. D, 58, 063001,
  \dodoi{10.1103/PhysRevD.58.063001}

\bibitem[{Jung \& Shin(2017)}]{Jung:2017flg}
Jung, S., \& Shin, C.~S. 2017.
\newblock \doarXiv{1712.01396}

\bibitem[{Liao {et~al.}(2019)Liao, Biesiada, \& Fan}]{Liao:2019aqq}
Liao, K., Biesiada, M., \& Fan, X.-L. 2019, Astrophys. J., 875, 139,
  \dodoi{10.3847/1538-4357/ab1087}

\bibitem[{{Manchester} {et~al.}(2005){Manchester}, {Hobbs}, {Teoh}, \&
  {Hobbs}}]{ATNF}
{Manchester}, R.~N., {Hobbs}, G.~B., {Teoh}, A., \& {Hobbs}, M. 2005, \aj, 129,
  1993, \dodoi{10.1086/428488}

\bibitem[{{Marchant} {et~al.}(2020){Marchant}, {Breivik}, {Berry}, {Mandel}, \&
  {Larson}}]{2020PhRvD.101b4039M}
{Marchant}, P., {Breivik}, K., {Berry}, C. P.~L., {Mandel}, I., \& {Larson},
  S.~L. 2020, \prd, 101, 024039, \dodoi{10.1103/PhysRevD.101.024039}

\bibitem[{Ng {et~al.}(2018)Ng, Wong, Broadhurst, \& Li}]{Ng2018}
Ng, K. K.~Y., Wong, K. W.~K., Broadhurst, T., \& Li, T. G.~F. 2018, Phys. Rev.
  D, 97, 023012, \dodoi{10.1103/PhysRevD.97.023012}

\bibitem[{{Paczynski}(1990)}]{Paczynski:1990}
{Paczynski}, B. 1990, \apj, 348, 485, \dodoi{10.1086/168257}

\bibitem[{Punturo {et~al.}(2010)Punturo, Abernathy, Acernese, Allen, Andersson,
  Arun, Barone, Barr, Barsuglia, Beker, Beveridge, Birindelli, Bose, Bosi,
  Braccini, Bradaschia, Bulik, Calloni, Cella, Mottin, Chelkowski, Chincarini,
  Clark, Coccia, Colacino, Colas, Cumming, Cunningham, Cuoco, Danilishin,
  Danzmann, Luca, Salvo, Dent, Derosa, Fiore, Virgilio, Doets, Fafone, Falferi,
  Flaminio, Franc, Frasconi, Freise, Fulda, Gair, Gemme, Gennai, Giazotto,
  Glampedakis, Granata, Grote, Guidi, Hammond, Hannam, Harms, Heinert, Hendry,
  Heng, Hennes, Hild, Hough, Husa, Huttner, Jones, Khalili, Kokeyama, Kokkotas,
  Krishnan, Lorenzini, Lück, Majorana, Mandel, Mandic, Martin, Michel,
  Minenkov, Morgado, Mosca, Mours, Müller-Ebhardt, Murray, Nawrodt, Nelson,
  Oshaughnessy, Ott, Palomba, Paoli, Parguez, Pasqualetti, Passaquieti,
  Passuello, Pinard, Poggiani, Popolizio, Prato, Puppo, Rabeling, Rapagnani,
  Read, Regimbau, Rehbein, Reid, Rezzolla, Ricci, Richard, Rocchi, Rowan,
  Rüdiger, Sassolas, Sathyaprakash, Schnabel, Schwarz, Seidel, Sintes, Somiya,
  Speirits, Strain, Strigin, Sutton, Tarabrin, van~den Brand, van Leewen, van
  Veggel, van~den Broeck, Vecchio, Veitch, Vetrano, Vicere, Vyatchanin, Willke,
  Woan, Wolfango, \& Yamamoto}]{Punturo_2010}
Punturo, M., Abernathy, M., Acernese, F., {et~al.} 2010, Classical and Quantum
  Gravity, 27, 084007, \dodoi{10.1088/0264-9381/27/8/084007}

\bibitem[{{PyCBC}(2022)}]{pycbcpsd}
{PyCBC}. 2022, {PyCBC} {PSD} Package.
\newblock \url{https://pycbc.org/pycbc/latest/html/pycbc.psd.html}

\bibitem[{Reed {et~al.}(2021)Reed, Deibel, \& Horowitz}]{Reed:2021scb}
Reed, B.~T., Deibel, A., \& Horowitz, C.~J. 2021, Astrophys. J., 921, 89,
  \dodoi{10.3847/1538-4357/ac1c04}

\bibitem[{Sartore {et~al.}(2010)Sartore, Ripamonti, Treves, \&
  Turolla}]{Sartore:2010}
Sartore, N., Ripamonti, E., Treves, A., \& Turolla, R. 2010, Astronomy \&
  Astrophysics, 510, A23

\bibitem[{Sch{\"o}del {et~al.}(2002)Sch{\"o}del, Ott, Genzel, Hofmann, Lehnert,
  Eckart, Mouawad, Alexander, Reid, Lenzen, {et~al.}}]{schodel2002star}
Sch{\"o}del, R., Ott, T., Genzel, R., {et~al.} 2002, Nature, 419, 694

\bibitem[{{Smith} {et~al.}(2019){Smith}, {Bianconi}, {Jauzac}, {Richard},
  {Robertson}, {Berry}, {Massey}, {Sharon}, {Farr}, \& {Veitch}}]{Smith2019}
{Smith}, G.~P., {Bianconi}, M., {Jauzac}, M., {et~al.} 2019, \mnras, 485, 5180,
  \dodoi{10.1093/mnras/stz675}

\bibitem[{Suvorov(2021)}]{Suvorov:2021uvd}
Suvorov, A.~G. 2021.
\newblock \doarXiv{2112.01670}

\bibitem[{Treves {et~al.}(2000)Treves, Turolla, Zane, \& Colpi}]{Treves:1999ne}
Treves, A., Turolla, R., Zane, S., \& Colpi, M. 2000, Publ. Astron. Soc. Pac.,
  112, 297, \dodoi{10.1086/316529}

\bibitem[{Urrutia \& Vaskonen(2021)}]{Urrutia:2021qak}
Urrutia, J., \& Vaskonen, V. 2021.
\newblock \doarXiv{2109.03213}

\end{thebibliography}
\end{document}